\documentclass[%
 preprint,
 amsmath,amssymb,
 aps, physrev,
]{revtex4-2}

\usepackage{graphicx}% Include figure files
\usepackage{dcolumn}% Align table columns on decimal point
\usepackage{bm}% bold math
\usepackage[dvipsnames]{xcolor}
\usepackage{url}
\usepackage{xcolor}
\usepackage{amsmath}
\usepackage[normalem]{ulem}

\begin{document}

\preprint{Preprint}

\title{\textbf{Method for Determining the Parameters of a Ring-like Structure from the Visibility Function Shape} 
}% 

\author{S.~V.~Chernov}
\email{Contact author: chernov@td.lpi.ru}
\affiliation{Astro Space Center, Lebedev Physical Institute, Russian Academy of Sciences, Profsoyuznaya str. 84/32, Moscow, 117997, Russian Federation}%
\author{M.~A.~Shchurov}%
\affiliation{Astro Space Center, Lebedev Physical Institute, Russian Academy of Sciences, Profsoyuznaya str. 84/32, Moscow, 117997, Russian Federation}
\author{I.~I.~Bulygin}
\affiliation{Astro Space Center, Lebedev Physical Institute, Russian Academy of Sciences, Profsoyuznaya str. 84/32, Moscow, 117997, Russian Federation}
\author{A.~G.~Rudnitskiy}
\affiliation{Astro Space Center, Lebedev Physical Institute, Russian Academy of Sciences, Profsoyuznaya str. 84/32, Moscow, 117997, Russian Federation}

\date{\today}

\begin{abstract}
Black hole images obtained by very long baseline interferometry (VLBI) by the Event Horizon Telescope are a new tool for testing general relativity in super-strong gravitational fields. These images demonstrated a ring-like structure which can be explained as the black hole shadow image. To date, there are no reliable methods for determining the parameters of these ring-like structures, such as diameter, width, and asymmetry. In this paper, an algorithm for determining black hole image parameters is proposed using a Gaussian asymmetric ring as an example. Using the proposed method, the diameter and asymmetry parameters of the image of a supermassive black hole in the galaxy M87$^{*}$ were estimated based on observational data obtained by the Event Horizon Telescope group.
\end{abstract}

\maketitle

\section{\label{sec:intro}Introduction}

The Event Horizon Telescope (EHT) team has opened a new era in the study of the super-strong gravitational field by obtaining the first-ever images of two supermassive black holes (SMBH) in radio astronomy: M87$^{*}$ \cite{EHT_2019} and Sgr~A$^{*}$ \cite{EHT_2022}. The resulting images revealed a ring-like structure, with a dark spot -- the black hole shadow. Based on these images, it was possible to estimate the mass of these supermassive black holes \cite{EHT_2019_VI, EHT_2022_IV}. However, due to insufficient angular resolution and limited $(u,v)$ coverage, it was not possible to determine the spin of the black holes. In order to determine the spin of a black hole, not only observations with higher angular resolution and better $(u,v)$ coverage are required, but also new efficient algorithms for analyzing the obtained data.
% предложил бы переформулировать как:
%
% not only observations... but new algorithms are required.
%
% Потому что мы говорим про актуальность не просто всей темы с черными дырами, но именно нашей части работы внутри темы.
% (с) Игорь

The EHT observations of M87$^*$ and Sgr~A$^*$ used ground millimeter telescopes capable of operating at a wavelength of 1.3~mm. With the acquisition of images of these supermassive black holes, it became clear that ground-based instruments are not able of obtaining images with a higher angular resolution and high-quality $(u,v)$ coverage. For these purposes, the ``Millimetron'' observatory is currently under development, which will be capable of conducting space-ground radio interferometric (VLBI) observations in the wavelength range of 0.7 -- 7~mm \cite{Novikov2021}. Also, various concepts of VLBI interferometers consisting solely of space radio telescopes are being considered \cite{Hong2014, Roelofs2019, Gurvits2021, Kudriashov2021a, Kudriashov2021b, Kurczynski2022, Trippe2023, Rudnitskiy2023}.

In VLBI observations, the main source of information is the complex visibility function $V$. By studying its distribution, it is possible to estimate the SMBH parameters. Earlier, in the work \cite{Johnson2020}, the main provisions on the expected form of the visibility function for observed images of black holes were considered. It is critical to develop reliable methods for determining SMBH parameters from detailed radio images. This is an urgent task both for the study of already obtained data and for the formation of tools for analyzing observations in future missions aimed at a detailed study of the SMBH close vicinity by VLBI methods \cite{Novikov2021, Kudriashov2021a, Kurczynski2022, Rudnitskiy2023}.

In this paper, we present an analytical method for determining the parameters of a ring-like structure, including its diameter, width, and asymmetry parameters. These algorithms can be further used to estimate the spin of a black hole. The method is based on a simple model of the SMBH silhouette based on a ring-like asymmetric brightness distribution. This structure is described by an asymmetric Gaussian brightness distribution, which allows us to obtain an approximate analytical expression for the visibility function. This approach, in turn, improves the accuracy of approximation of simulated and real observations. Using the above simplified model and the proposed method, we estimated the diameter, width, and asymmetry of the ring-like structure from real EHT observations of M87$^{*}$.

\section{Analytical Models}
An idealized model of a ring-like structure with a non-uniform and asymmetric brightness distribution was used for subsequent evaluation of the ring-like structure parameters. Such a structure can be interpreted as an image of a black hole \cite{EHT_2019,EHT_2022} and the main objective of this paper is to obtain the ring parameters (diameter, width, and asymmetry parameter). To do this, it is first necessary to consider a ring-like structure in polar coordinates $r,\phi_r$. Its brightness can be represented as follows:

\begin{equation}
 I(r,\phi_r)=I_r(r)I_\phi(\phi_r),
 \label{brightness}
\end{equation}

where $I_r$ is a function that describes the non-uniform distribution of the ring-like structure brightness along the radius, and $I_\phi$ is a function that describes the asymmetric distribution of the ring-like structure brightness by angle. It is assumed that the ring-like structure itself has the shape of a circle with a zero eccentricity. The circle center is located at $x_0=0$, $y_0=0$. For convenience, the ring's integral brightness is normalized to one Jy. All the coordinates on the picture like $x, y, r$ are the angular coordinates on the celestial sphere and are typically measures in $\mu$as.

The asymmetric brightness distribution can be defined as follows \cite{Tiede2022_A}:

\begin{equation}
 I_\phi(\phi_r)=\left(1-B\cdot\sin^2\frac{\phi_r-\phi_0}{2}\right)^n,
 \label{axibrightness}
\end{equation}

where $\phi_0$ is the direction to the maximum brightness of the ring, $n$ is the degree of asymmetry. The coefficient $B$ is the asymmetry parameter and takes values in the range $|B|\leq1$. In reality, the asymmetry parameter depends on many variables. In particular, on the spin of the black hole and the angle of inclination of the observer to the black hole's rotation axis. In this paper, the case is considered when the asymmetry parameter is constant. Additionally, the case where the degree of asymmetry is $n=1$ is also considered.

The non-uniform brightness distribution along the radius can be described by a Gaussian function:
\begin{equation}
 I_r(r)=e^{-(\frac{r-r_0}{\Delta r})^2},
 \label{rbrightness}
\end{equation}

where $r_0$ is the radius of the ring, $\Delta r$ is the thickness of the ring. This brightness distribution (\ref{brightness}) is asymmetric and non-uniform, i.e. it depends on both $r$ and $\phi_r$. Comparing with exact analytical expressions, it is also necessary to consider the case of a homogeneous asymmetric ring brightness distribution $I\sim I_\phi(\phi_r)\delta(r-r_0)$, where $\delta$ is the delta function \cite{Johnson2020}.

\begin{figure}
\centering
\includegraphics[width=0.75\linewidth]{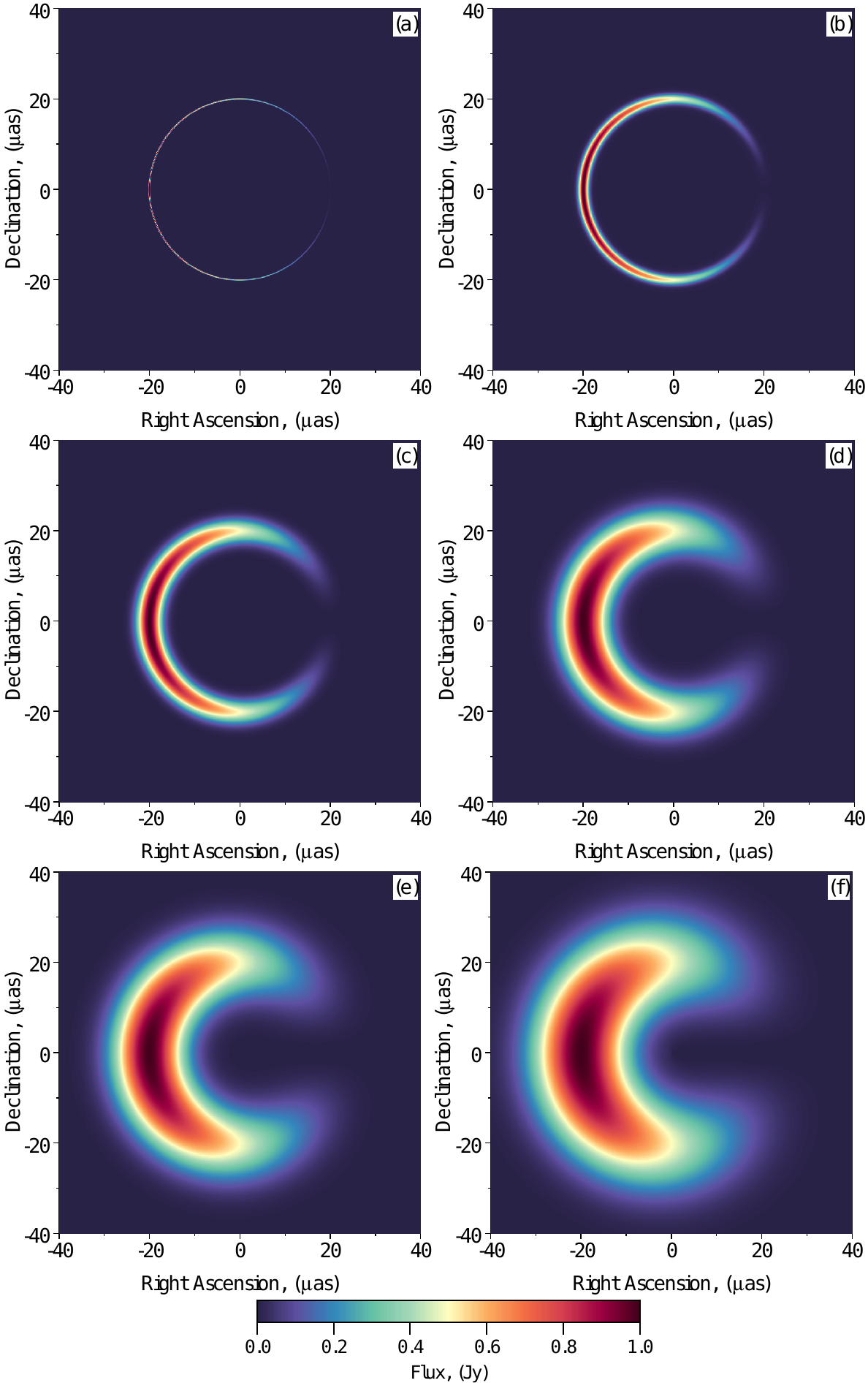}
\caption{Brightness distribution in the ring depending on the ring width $\Delta r$ for the asymmetry parameter $B=1$, the degree of asymmetry $n=1$ and the ring radius $r_0=20$ $\mu$as. The ring width $\Delta r$ is: 0.1 (a), 1 (b), 2.5 (c), 5 (d), 7.5 (e) and 10 (f) $\mu$as.}
\label{fig1}
\end{figure}

Examples of brightness distribution in rings are shown in Fig.~\ref{fig1} for the following parameters: ring radius $r_0=20$ $\mu$as, asymmetry parameter $B=1$ and ring width $\Delta r=0.1;1.0;2.5;5.0;7.5;10.0$ $\mu$as.

Visibility function $V$ is the interferometer response to the signal distribution in the sky \cite{Thompson2017}. In this paper, the ring parameters are determined from the shape of the visibility function distribution depending on the interferometer baseline projection.

Studying the shape of the two-dimensional visibility function and comparing the theory with numerical observations allows us to estimate the ring parameters. The visibility function is complex and depends on two coordinates. In polar coordinates $u,\phi_u$ it can be represented as \cite{Thompson2017}:

\begin{equation}
 V(u,\phi_u)=\int\int I(r,\phi_r) e^{-2\pi iur\cos(\phi_r-\phi_u)}rdrd\phi_r,
 \label{vid}
\end{equation}

where $u$ is the dimensionless interferometer baseline projection in units of wavelengths, $r$ is the dimensionless image size in radians.

Substituting the expression (\ref{brightness}) with an asymmetric brightness distribution (\ref{axibrightness}) into (\ref{vid}), the integral over $\phi_r$ can be calculated directly. As a result,

\begin{equation}
 V(u,\phi_u)=\pi\int [(2-B)J_0(2\pi ur)-iBJ_1(2\pi ur)\cos(\phi_u-\phi_0)]I_r(r)rdr,
 \label{V1}
\end{equation}

where $J_0$ and $J_1$ are the Bessel functions of the first kind of zero and first order, respectively. From this expression it is clear that the visibility function is complex due to the ring brightness asymmetry. Consequently, the visibility function modulus will differ significantly in two perpendicular directions at $\phi_u=0$ and $\phi_u=\pi/2$. Below are several cases.

\subsection{Infinitely thin homogeneous ring}
For the case of an infinitely thin homogeneous ring \cite{Johnson2020}, when

\begin{equation}
I_r(r)=\frac{\delta(r-r_0)}{2\pi r_0},
\end{equation}

as a result of integration the visibility function is obtained:

\begin{equation}
 V(u,\phi_u)=\frac{1}{2}\bigg[(2-B)J_0(2\pi ur_0)-iBJ_1(2\pi ur_0)\cos(\phi_u-\phi_0)\bigg].
 \label{Vaxithinring}
\end{equation}

This visibility function is complex due to the ring brightness asymmetry. In the case of a symmetric ring brightness distribution ($B=0$), this solution becomes as represented in \cite{Johnson2020}.

\subsection{Thick homogeneous ring}
For a thick homogeneous ring, when $I_r(r)=1$, the integral (\ref{V1}) can be calculated directly \cite{Chernov2021}. As a result,

\begin{eqnarray}
 V(u,\phi_u)=\frac{1}{2u}\bigg[(2-B)(bJ_1(2\pi ub)-aJ_1(2\pi ua))-\nonumber\\
 -i\frac{\pi}{2} B\cos(\phi_u-\phi_0)[b(J_1(2\pi ub)H_0(2\pi ub)-J_0(2\pi ub)H_1(2\pi ub))-\nonumber\\
 -a(J_1(2\pi ua)H_0(2\pi ua)-J_0(2\pi ua)H_1(2\pi ua))]\bigg],
 \label{V2}
\end{eqnarray}

where $b$ is the outer radius of the ring, $a$ is the inner radius of the ring, $H_0$ and $H_1$ are the Struve functions. This visibility function is also complex due to the ring brightness asymmetry.

\subsection{Parabolic distribution of brightness in a ring}

It is pertinent to consider the case of a parabolic brightness distribution \cite{Chernov2021}:

\begin{eqnarray}
 I_r(r)=\frac{1}{\pi dw}\bigg[1-\left(\frac{2}{w}\right)^2\left(r-\frac{d}{2}\right)^2\bigg],
\end{eqnarray}

where $d=a+b$ is the diameter, and $w=b-a$ is the width of the ring. For such a brightness distribution, the visibility function has a more complex form:

\begin{eqnarray}
V=\frac{2-B}{dw}\bigg[\left(1-\frac{d^2}{w^2}\right)\left(\frac{b}{2\pi u}J_1(2\pi ub)-\frac{a}{2\pi u}J_1(2\pi ua)\right)+\nonumber\\
+\frac{2d}{\pi uw^2}\bigg(b^2J_1(2\pi ub)-a^2J_1(2\pi ua)+
\frac{b}{4u}(J_0(2\pi ub)H_1(2\pi ub)-J_1(2\pi ub)H_0(2\pi ub))-\nonumber\\
-\frac{a}{4u}(J_0(2\pi ua)H_1(2\pi ua)-J_1(2\pi ua)H_0(2\pi ua))\bigg)-\nonumber\\
-\frac{2}{\pi^2 u^2w^2}\bigg(b^2(J_2(2\pi ub)-\pi ub J_3(2\pi ub))-
a^2(J_2(2\pi ua)-\pi ua J_3(2\pi ua))\bigg)\bigg]-\nonumber\\
-i\frac{B}{dw}\cos(\phi_u-\phi_0)\bigg[\left(1-\frac{d^2}{w^2}\right)\bigg(\frac{b}{4u}(J_1(2\pi ub)H_0(2\pi ub)-H_1(2\pi ub)J_0(2\pi ub))-\nonumber\\
-\frac{a}{4u}(J_1(2\pi ua)H_0(2\pi ua)-H_1(2\pi ua)J_0(2\pi ua))\bigg)+\nonumber\\
+\frac{2d}{\pi uw^2}\left(b^2J_2(2\pi ub)-a^2J_2(2\pi ua)\right)-\nonumber\\
-\frac{1}{w^2\pi^2u^2}\bigg(\frac{3b}{4u}(J_0(2\pi ub)H_1(2\pi ub)-J_1(2\pi ub)H_0(2\pi ub))-b^2(2\pi ubJ_0(2\pi ub)-3J_1(2\pi ub))-\nonumber\\
-\frac{3a}{4u}(J_0(2\pi ua)H_1(2\pi ua)-J_1(2\pi ua)H_0(2\pi ua))+a^2(2\pi uaJ_0(2\pi ua)-3J_1(2\pi ua))\bigg)\bigg]
\label{V-parabolic}
\end{eqnarray}

All the above solutions are analytically exact.

\subsection{Approximate expressions for the visibility function}

For a non-uniform brightness distribution (\ref{rbrightness}), the integral (\ref{V1}) can be approximated, for example, by the saddle-point method. As a result, the visibility function of will have the following form:

\begin{eqnarray}
 V(u,\phi_u)=\pi^{3/2}r_0\Delta r\bigg[(2-B)J_0(2\pi ur_0)-iBJ_1(2\pi ur_0)\cos(\phi_u-\phi_0)\bigg]
\end{eqnarray}

This solution is approximate, from which it is clear that the ring width cannot be determined explicitly. In order to determine it, one can assume that the ring radius is much larger than the ring width. In addition, the interferometer baseline projection must be large enough. The Appendix contains the visibility function for this case. This solution (\ref{Vthick}) for the visibility function, taking into account the brightness distribution (\ref{axibrightness}) and (\ref{rbrightness}), is also approximate and depends on both the ring radius and its width.

\section{Determining the Ring Parameters}
\subsection{Ring Diameter}
To estimate a ring diameter, it is necessary to know the position of the zeros of the Bessel function $J_0$, which are determined by the following expression: 

\begin{eqnarray}
j_{0,n}=\frac{\pi}{4}(4n+3)+\frac{1}{2\pi(4n+3)}-\frac{31}{6\pi^3(4n+3)^3}, (n=0,1,2,...),
\end{eqnarray}

where $n$ is the zero number. The first zero of the Bessel function $J_0$ is approximately equal to $j_{0,0}\approx2.40$, the second zero is $j_{0,1}\approx5.52$, the third zero is $j_{0,2}\approx8.65$, and so on. Therefore, the ring radius can be determined from $j_{0,n}=2\pi u_n r_0$, where $u_n$ are the visibility function zero positions.

Since the inverse Fourier transform, which is applied to the original model to obtain the visibility function, is a linear operation, the linear operations of shift and rotation can also be applied to the original model in such a way that the center of the ring coincides with the origin of coordinates, and the position of the intensity maximum on the ring coincides with the negative direction of the X-axis. This allows us to simplify the calculations and consider only two extreme mutually perpendicular sections of the two-dimensional visibility function along the X and Y axes.

\begin{figure}
\centering
\includegraphics[width=1\linewidth]{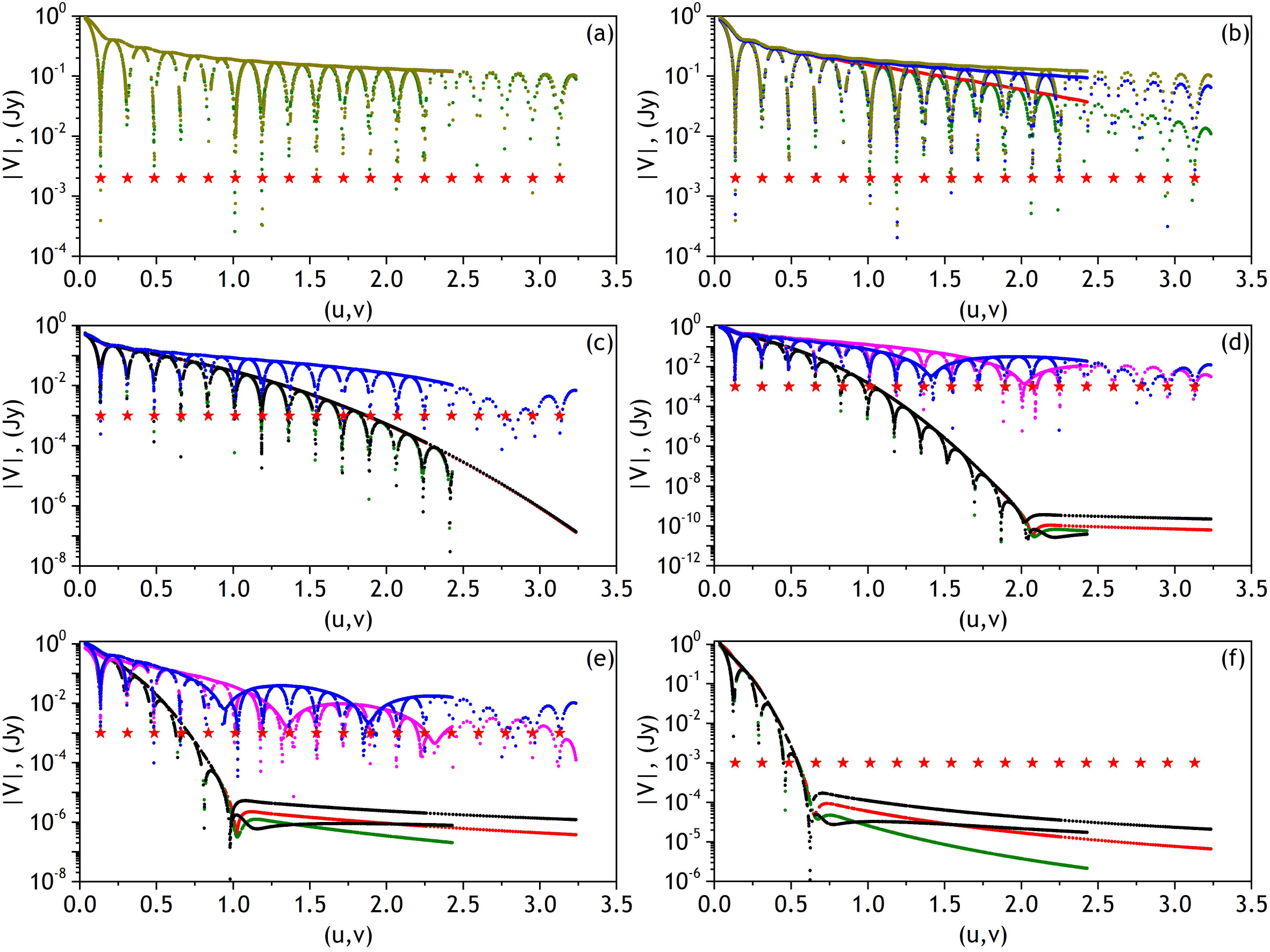}
\caption{Visibility functions in two perpendicular directions obtained for the images shown in Fig.~\ref{fig1}: (a) -- ring width $w=0.1$ $\mu$as, (b) -- ring width $w=1.0$ $\mu$as, (c) -- ring width $w=2.5$ $\mu$as, (d) -- ring width $w=5.0$ $\mu$as, (e) -- ring width $w=7.5$ $\mu$as, (f) -- ring width $w=10.0$ $\mu$as. Green dots correspond to the modulus of the visibility function $|V|$ in the $v$ direction, red ones in the $u$ direction, respectively.}
\label{fig2}
\end{figure}

Fig.~\ref{fig2} shows the distribution of the modulus of the visibility function in two perpendicular directions for the images presented in Fig.~\ref{fig1} for rings with widths: $w=0.1$ (a), $w=1.0$ (b), $w=2.5$ (c), $w=5.0$ (d), $w=7.5$ (e), $w=10.0$ (f) $\mu$as. In Fig.~\ref{fig2} (a), (b) yellow dots are analytical points obtained by ~(\ref{Vaxithinring}) for an infinitely thin ring, blue dots belong to analytical points of a thick disk (\ref{V2}), magenta dots are analytical points (\ref{V-parabolic}), black dots are analytical points (\ref{Vthick}). Green dots correspond to the modulus of the visibility function in the direction of $v$ ($\phi_u=\pi/2$), red dots in the direction of $u$ ($\phi_u=\pi$). Red stars correspond to the Bessel function $j_{0,n}$ zeroes. Comparing the position of these points with the position of the minimum of numerical observations, it is possible to estimate the ring radius.

As can be seen from Fig.~\ref{fig2} (a), the analytical approximation of the thin ring agrees with the numerical observations. For a ring with a width $w=1.0$ $\mu$as (see Fig~\ref{fig2} (b), the analytical thick ring describes the numerical observations better than the infinitely thin ring. For baselines greater than one Earth diameter, the deviations of the analytical formulas from the numerical observations become significant. As a result, the visibility function zeroes coincide in all cases. For $w=2.5$ $\mu$as (Fig~\ref{fig2}(c)) the visibility function of the infinitely thin ring is not shown here, due to a strong discrepancy between the numerical observations. The analytical thick ring agrees well with numerical observations for baselines smaller than $u<0.6$ Earth diameters. 

From Fig~\ref{fig2} (d) it is clear that the homogeneous and parabolic rings describe numerical observations accurately only for baselines smaller than $u<0.3$ Earth diameters. The zero of the parabolic ring envelope agrees well with the zero of the numerical observations envelope. The zero of the homogeneous ring diverges strongly from the zero of the numerical observations envelope. The approximate solution (\ref{Vapprox}) describes the numerical observations best. For the case shown in Fig.~\ref{fig2} (e), homogeneous and parabolic rings describe numerical observations accurately only for baseline projections of $u<0.2$ Earth diameters. The opposite case is observed in Fig.\ref{fig2} (d), when the zero of the homogeneous ring envelope coincides well with the zero of the numerical observations envelope, and the zero of the parabolic ring envelope diverges strongly from the zero of the numerical observations envelope. For the case $w=10.0$ $\mu$as (Fig.~\ref{fig2} (f)) the visibility function of both the homogeneous ring and the parabolic ring is not given due to strong deviations from the numerical observations.

The Table~\ref{tab1} presents the values of the zeroes of the Bessel function $j_{0,n}$ obtained from the visibility functions (Fig. (\ref{fig2})) and the estimated ring radii. From Table~\ref{tab1} it is clear that even for fairly wide rings this estimate determines the ring radii with reasonable accuracy.

\begin{table}
\centering
\caption{The first three visibility function zeros (in Earth diameters --- ED) and the ring diameter estimate based on these zeros in $\mu$as.}
\label{tab1}
\begin{tabular}{|c|c|c|c|c|c|c|}
\hline
 $w$, $\mu$as & 0.1 & 1.0 & 2.5 & 5.0 & 7.5 & 10.0\\
\hline
 $u_0$, ED & 0.135 & 0.135 & 0.134 & 0.133 & 0.130 & 0.126 \\
\hline
 $u_1$, ED & 0.309 & 0.308 & 0.307 & 0.306 & 0.306 & 0.289 \\
\hline
 $u_2$, ED & 0.485 & 0.484 & 0.482 & 0.483 & 0.464 & 0.462 \\
\hline
 $r_0$, $\mu$as & 20.0 & 20.0 & 20.1 & 20.2 & 20.7 & 21.4 \\
\hline
 $r_0$, $\mu$as & 20.0 & 20.1 & 20.2 & 20.2 & 20.2 & 21.4\\
\hline
 $r_0$, $\mu$as & 20.0 & 20.1 & 20.1 & 20.1 & 20.9 & 21.0\\
\hline
\end{tabular}
\end{table}

In turn, it should be noted that the approximate solution of a thick disk with a Gaussian distribution (\ref{Vthick}) describes the numerical observations for all ring widths (black dots) quite accurately. Therefore, by fitting the numerical observations using this model, it is possible to estimate the ring thickness.

\subsection{Ring thickness}
The effect of oscillations of two periodic functions is used to estimate ring thickness. For simplicity, a homogeneous ring with width $w=b-a$ and visibility function (\ref{V2}) are considered. In the direction of $\phi_u=\pi/2$, the visibility function is determined by the difference between the first-order Bessel functions:

\begin{eqnarray}
 V(u,\phi_u=\pi/2)=\frac{2-B}{2u}(bJ_1(2\pi ub)-aJ_1(2\pi ua)).
\end{eqnarray}

Here it is necessary to consider the case when the interferometer baseline projection is much larger than the Earth diameter, i.e. $2\pi ua\gg1$. In this case, expanding the Bessel functions in series, we obtain:

\begin{eqnarray}
 V(u,\phi_u=\pi/2)\approx\frac{(2-B) r_0^{1/2}}{\pi u^{3/2}}\sin\left(\frac{\pi}{4}+\pi u(a+b)\right)\sin(\pi uw).
\end{eqnarray}

Low-frequency oscillations with spatial frequency $w$ are the envelope for high-frequency oscillations with spatial frequency $(a+b)$. In order to calculate the ring thickness, it is necessary to determine the envelope zeros. To do this, set $\sin(\pi uw)=0$. It follows that, $w\approx1/u$. Thus, the ring thickness can be estimated from the zero location in the envelope shape. For rings with widths $w=0.1;1.0;2.5$, the envelope is approximately zero at $u,v\approx71;7.1;2.8$, which significantly exceeds the maximum baseline projection of the considered interferometer configuration. This means that it is impossible to estimate the ring width of these cases. For the remaining rings ($w=5.0,7.5,10.0$), the first zeros in the envelope $u_0$ and the corresponding estimates of the ring width $w_1$ are presented in Table~\ref{tab2}.

\begin{table}
\caption{The first zeros in the envelope $u_{0}$ and the estimate of the ring width $w_1$ for $w=5.0,7.5,10.0$ $\mu$as.}
\label{tab2}
\centering
\begin{tabular}{|c|c|c|c|c|}
\hline
 $w$, $\mu$as  & 5.0 & 7.5 & 10.0\\
\hline
 $u_0$, ED  & 2.07 & 1.02 & 0.67 \\
\hline
 $w_1$, ED  & 3.41 & 6.91 & 10.5 \\
\hline
\end{tabular}
\end{table}

\section{Visibility functions of two rings}

Consider cases with two thin rings of close radii and a combination of a thin and a thick ring. To study these cases, numerical observations were used with the space interferometer configuration described in \cite{Rudnitskiy2023}. The configuration of such an interferometer allows a relatively good $(u,v)$ coverage, but the maximum baseline projection for such a configuration does not exceed $\approx3.4$ Earth diameters.

\subsection{Two thin rings of close radii}

Let us consider the case when the image of a black hole consists of two thin rings with different, but very close radii ($r_1\approx r_2$). Such a configuration is possible at sufficiently high angular resolution. This is when the first thin ring corresponds to the first photon ring of the black hole, and the second to the second photon ring. The positions of the maximum brightness in the photon rings will depend on the spin of the black hole and the angle between the observer and the axis of rotation of the black hole. If the spin is zero, then the position of the brightness maxima of each subsequent ring will be shifted by the angle $\phi=\pi$. In the case of a non-zero spin and zero angle between the observer and the black hole's axis of rotation, the position of the maximum brightness can be found by the methods described in \cite{Andrianov2022}. Here we will consider several special cases.

As shown above, thin rings are well described by the visibility function of type (\ref{Vaxithinring}). Therefore, this visibility function will be used to define the formulas by which we will determine the radius and width of the rings. The approximate visibility function of two thin rings will be determined by the sum of the visibility functions of each ring (\ref{Vaxithinring}):

\begin{eqnarray}
V(u,\phi_u)=\frac{1}{2}\bigg[(2-B_1)J_0(2\pi ur_1)-iB_1J_1(2\pi ur_1)\cos(\phi_u-\phi_1)+\nonumber\\
+(2-B_2)J_0(2\pi ur_2)-iB_2J_1(2\pi ur_2)\cos(\phi_u-\phi_2)\bigg],
\end{eqnarray}

where $r_1$, $r_2$ are the radii of the first and second rings, $\phi_1$, $\phi_2$ are the positions of the maximum brightness of the rings, and $B_1$, $B_2$ is the asymmetry parameter of the rings.

\begin{figure}
\centering
\includegraphics[width=1\linewidth]{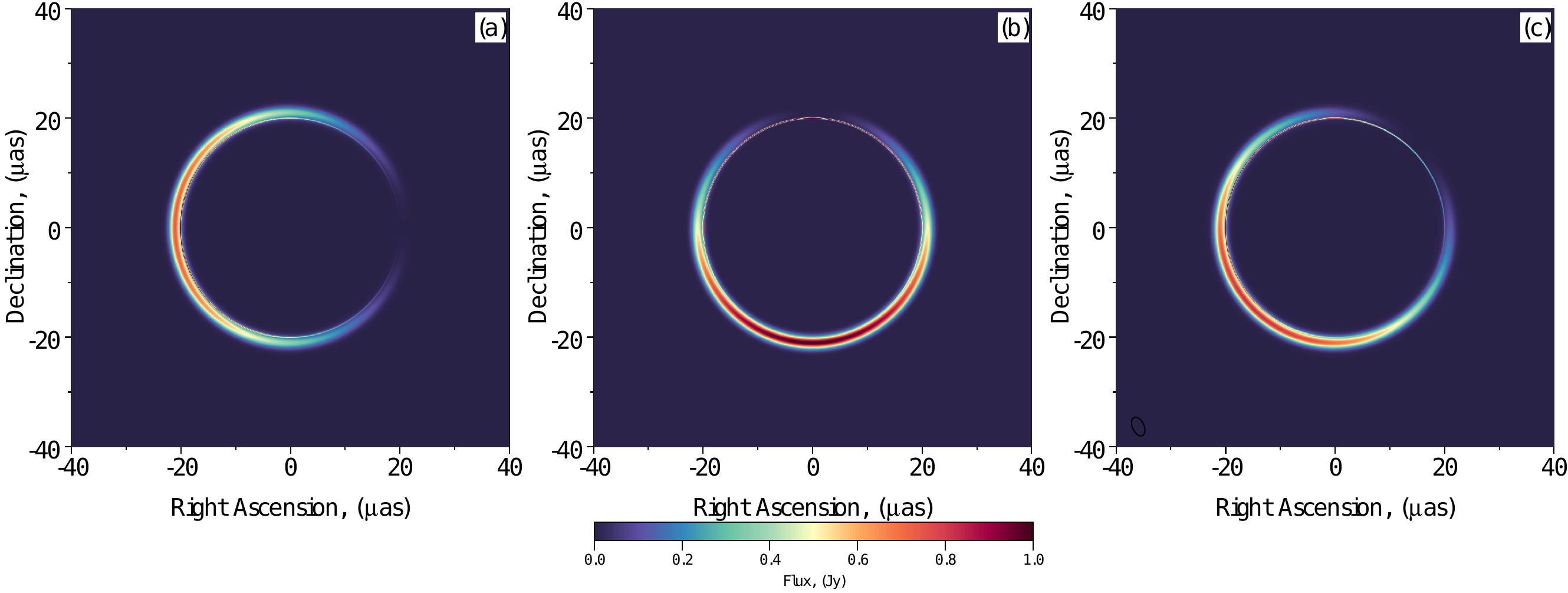}
\caption{An example of images with two thin rings with different maximum brightness positions. The inner ring radius is $r_1=20.0$ $\mu$as, the width is $w=0.1$ $\mu$as. The outer ring radius is $r_2=21.0$ $\mu$as, the width is $w=1.0$ $\mu$as. The maximum brightness is located at   $\phi_1=\phi_2=\pi$ (a), $\phi_1=\pi/2$, $\phi_2=3\pi/2$ (b), $\phi_1=3\pi/4$, $\phi_2=5\pi/4$ (c).}
\label{fig5}
\end{figure}

If the maxima of the brightness position are located in one direction, for example $\phi_1=\phi_2=\pi$ and have the same asymmetry parameter equal to unity ($B_1=B_2=1$), as shown in Fig.~\ref{fig5} (a), then the radii of the rings can be determined from the location of the zeros in two perpendicular directions. The visibility function in the direction $\phi_u=\pi/2$ will be equal to:

\begin{eqnarray}
%V=J_0(2\pi ur_1)+J_0(2\pi ur_2)\approx\frac{\cos(2\pi ur_1)+\sin(2\pi ur_1)+
%\cos(2\pi ur_2)+\sin(2\pi ur_2)}{\pi\sqrt{2ur_1}}=\\
%=\frac{\sin\left(2\pi ur_1+\frac{\pi}{4}\right)+\sin\left(2\pi ur_2+\frac{\pi}{4}\right)}{\pi\sqrt{ur_1}}=\frac{2\sin\left(\pi u(r_1+r_2)+\frac{\pi}{4}\right)\cos(\pi u(r_2-r_1))}{\pi\sqrt{ur_1}}
V=\frac{1}{2}(J_0(2\pi ur_1)+J_0(2\pi ur_2))
\approx\frac{\sin\left(\pi u(r_1+r_2)+\frac{\pi}{4}\right)\cos(\pi u(r_2-r_1))}{\pi\sqrt{ur_1}}.
\label{V2ringphipiphipi}
\end{eqnarray}

The visibility function in the perpendicular direction, $\phi_u=\pi$, will be equal to:

\begin{eqnarray}
V=\frac{1}{2}\sqrt{(J_0(2\pi ur_1)+J_0(2\pi ur_2))^2+(J_1(2\pi ur_1)+J_1(2\pi ur_2))^2}
\approx\nonumber\\
\approx\frac{\cos(\pi u(r_2-r_1))}{\pi\sqrt{ur_1}}.
\end{eqnarray}

It is also assumed here that the baseline projection is large enough ($2\pi ur_{1,2}\gg1$). The first zeros of the rapidly oscillating function are determined by the relations:

\begin{eqnarray}
u(r_1+r_2)=n-\frac{1}{4}.
\label{r1+r2}
\end{eqnarray}

The next relation is determined by the zeros of the envelope:

\begin{eqnarray}
u(r_2-r_1)=n+\frac{1}{2}. 
\label{r2-r1}
\end{eqnarray}

\begin{figure}
\centering
\includegraphics[width=0.75\linewidth]{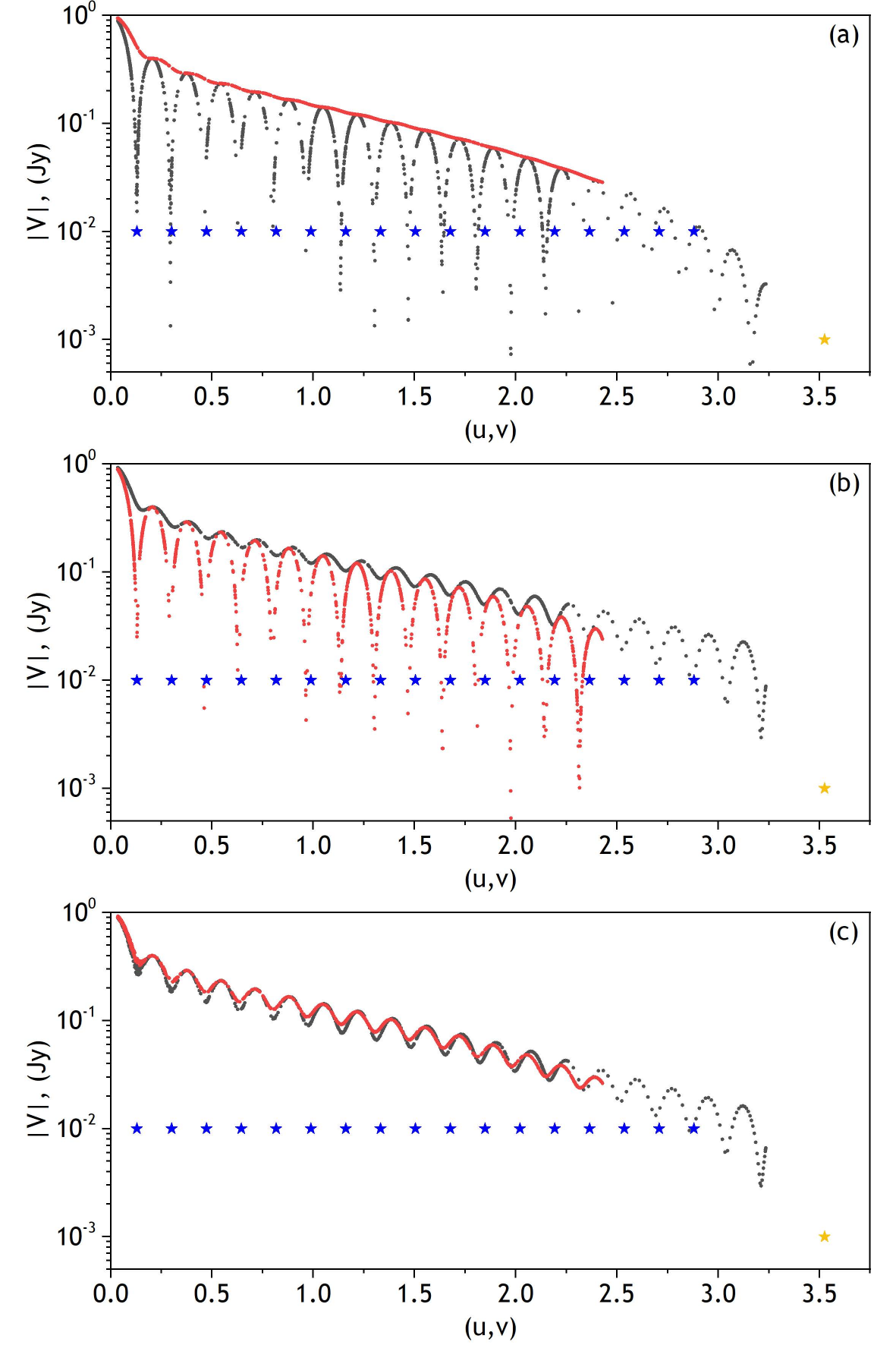}
\caption{Visibility functions for images consisting of two thin rings. Blue stars show the position of the visibility function zeros obtained from (\ref{r1+r2}). Orange circle shows the position of the visibility function zero obtaoined from (\ref{r2-r1}).}
\label{fig6}
\end{figure}

From these two relations the radius of each ring can be determined. The visibility function in two perpendicular directions is shown in Fig~\ref{fig6} (a). The red dots correspond to the direction, $\phi_u=\pi$, the black dots -- $\phi_u=\pi$. The first condition (\ref{r1+r2}) gives zero at $u=0.130$. From this we obtain $r_1+r_2=40.68$. The baseline projection is insufficient to determine the second condition (\ref{r2-r1}). Approximately, zero in the visibility function envelope is $u\approx3.5$.

If the positions of the brightness maxima are in opposite directions, for example $\phi_1=\frac{\pi}{2}$, $\phi_2=\frac{3\pi}{2}$, and have the same asymmetry parameter $B_1=B_2=1$, as shown in Fig.~\ref{fig5} (b), then the visibility function in the direction $\phi_u=\pi$ will be similar to (\ref{V2ringphipiphipi}), and in the direction $\phi_u=\pi/2$ will be equal to:

\begin{eqnarray}
V=\frac{1}{2}\sqrt{(J_0(2\pi ur_1)+J_0(2\pi ur_2))^2+(J_1(2\pi ur_1)-J_1(2\pi ur_2))^2}\approx\nonumber\\
%=\frac{1}{\pi\sqrt{ur_1}}\sqrt{\sin^2(\pi u(r_1+r_2)+\frac{\pi}{4})\cos^2(\pi u(r_2-r_1))+
%\sin^2\pi u(r_1-r_2)\cos^2(\pi u(r_1+r_2)-\frac{\pi}{4})}=\\
\approx\frac{\sin(\pi u(r_1+r_2)+\frac{\pi}{4})}{\pi\sqrt{ur_1}}.
\end{eqnarray}

Comparing the visibility function zeros with the previous expressions, we can conclude that the visibility function zeros are determined by the same expressions: $(\ref{r1+r2})$ and $(\ref{r2-r1})$. Fig.~\ref{fig6} (b) shows the visibility function in two perpendicular directions.

Here, the red dots correspond to $\phi_u=\pi$, the black dots to $\phi_u=\pi$. The first condition (\ref{r1+r2}) gives zero at $u=0.130$. From here we obtain $r_1+r_2=40.68$. The baseline projection is insufficient to determine the second condition (\ref{r2-r1}).

If the brightness maxima are located in perpendicular directions, for example $\phi_1=3\pi/4$, $\phi_2=5\pi/4$, and have the same asymmetry parameter $B_1=B_2=1$, as shown in the figure \ref{fig5} (c), then the visibility function in the direction $\phi_u=\pi$ will be equal to:

\begin{eqnarray}
V=\frac{1}{2}\sqrt{(J_0(2\pi ur_1)+J_0(2\pi ur_2))^2+\frac{1}{2}(J_1(2\pi ur_1)+J_1(2\pi ur_2))^2}\approx\nonumber\\
\approx\frac{\cos(\pi u(r_2-r_1))}{\pi\sqrt{ur_1}}\sqrt{\frac{1}{2}+\frac{1}{2}\cos^2(\pi u(r_1+r_2)-\frac{\pi}{4})}.
\end{eqnarray}

In the direction $\phi_u=\pi/2$ the visibility function is equal to:

\begin{eqnarray}
V=\frac{1}{2}\sqrt{(J_0(2\pi ur_1)+J_0(2\pi ur_2))^2+\frac{1}{2}(J_1(2\pi ur_1)-J_1(2\pi ur_2))^2}\approx\nonumber\\
\approx\frac{\sin(\pi u(r_2+r_1)+\frac{\pi}{4})}{\pi\sqrt{ur_1}}\sqrt{\frac{1}{2}+\frac{1}{2}\cos^2(\pi u(r_1-r_2))}.
\end{eqnarray}

As in the previous cases, the visibility function zeros are determined by (\ref{r1+r2}) and (\ref{r2-r1}). Fig.~\ref{fig6} (c) shows the visibility function in two perpendicular directions. The first condition (\ref{r1+r2}) gives zero at $u=0.135$. From here we obtain $r_1+r_2=39.17$. The baseline projection is insufficient to determine the second condition (\ref{r2-r1}).

From the above, it can be concluded that in order to determine the radii of two thin rings, it is necessary to find the zeros of the visibility function in two perpendicular directions and use the expressions (\ref{r1+r2}) and (\ref{r2-r1}). Due to insufficient angular resolution, it is impossible to determine the envelope zero from numerical observations. The maximum baseline projection is $u\approx3.4$, while according to theory the envelope zero should be located around $u\approx3.6$.

\subsection{Thick and thin ring}

A thick ring containing a thin ring may be observed in Event Horizon Telescope observations. However, due to a lack of angular resolution, the thick ring cannot be resolved. For a more visual illustration and to simplify analytical calculations, we consider not the modulus of the visibility function, but separately its real and imaginary parts. We assume that the asymmetry parameter is equal to one for two rings $B_1=B_2=1$. The brightness of each ring at maximum was set to 0.5 Jy. The resulting visibility function will be determined by the expression:

\begin{eqnarray}
 V=\frac{1}{2}[J_0(2\pi ur_1)-iJ_1(2\pi ur_1)\cos(\phi_u-\phi_1)]+\nonumber\\
 +\frac{1}{2u}\bigg[(bJ_1(2\pi ub)-aJ_1(2\pi ua))-\nonumber\\
 -i\frac{\pi}{2} \cos(\phi_u-\phi_2)[b(J_1(2\pi ub)H_0(2\pi ub)-J_0(2\pi ub)H_1(2\pi ub))-\nonumber\\
 -a(J_1(2\pi ua)H_0(2\pi ua)-J_0(2\pi ua)H_1(2\pi ua))]\bigg].
\end{eqnarray}

The resulting visibility function consists of the visibility function of the thin (\ref{Vaxithinring}) and thick (\ref{V2}) rings.
Then the real part of the visibility function does not depend on the direction of the maximum brightness (from the angle $\phi_u$) and is given by the expression:

\begin{eqnarray}
 Re V=\frac{1}{2}J_0(2\pi ur_1)
 +\frac{1}{2u}(bJ_1(2\pi ub)-aJ_1(2\pi ua))\approx\nonumber\\
 \approx\frac{\sin(2\pi ur_1+\frac{\pi}{4})}{2\pi\sqrt{ur_1}}+\frac{\sqrt{a+b}}{\pi u\sqrt{2u}}\sin(\pi u(b-a))\cos\left(\pi u(a+b)-\frac{\pi}{4}\right),
\end{eqnarray}

where it was also assumed that the baseline projection is large enough $2\pi ur_1\sim2\pi ua\sim2\pi ub\gg1$. The imaginary part of the visibility function will depend on $\phi_u$, as well as on the direction of maximum brightness in each ring $\phi_1, \phi_2$. The imaginary part of the visibility function is:

\begin{eqnarray}
Im V=-\frac{1}{2}J_1(2\pi ur_1)\cos(\phi_u-\phi_1)-\nonumber\\
-\frac{\pi}{4u}
[b(J_1(2\pi ub)H_0(2\pi ub)-J_0(2\pi ub)H_1(2\pi ub))-\nonumber\\
 -a(J_1(2\pi ua)H_0(2\pi ua)-J_0(2\pi ua)H_1(2\pi ua))]\cos(\phi_u-\phi_2)\approx\nonumber\\
\approx-\frac{\sin(2\pi ur_1-\frac{\pi}{4})}{2\pi\sqrt{ur_1}}\cos(\phi_u-\phi_1)+\nonumber\\
+\frac{\sqrt{a+b}}{\pi u\sqrt{2u}}\sin(\pi u(b-a))\cos\left(\pi u(a+b)+\frac{\pi}{4}\right)\cos(\phi_u-\phi_2).
\end{eqnarray}

\begin{figure}
\centering
\includegraphics[width=1\linewidth]{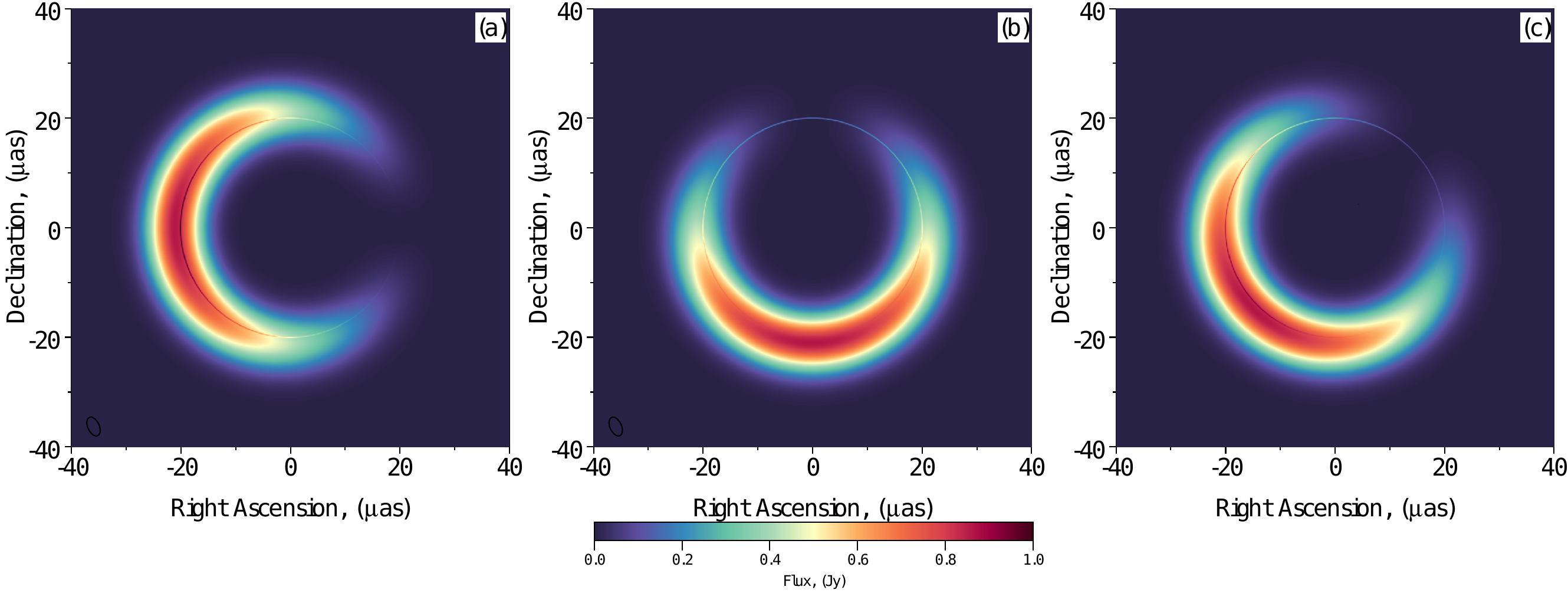}
\caption{An example of images consisting of one thick and one thin ring. The radius and width of the thin ring are $r_1=20.0$ and $w_1=0.1$ $\mu$as. The radius and width of the thick ring are $r_1=21.0$ and $w_1=5.0$ $\mu$as.}
\label{fig7}
\end{figure}

\begin{figure}
\centering
\includegraphics[width=0.75\linewidth]{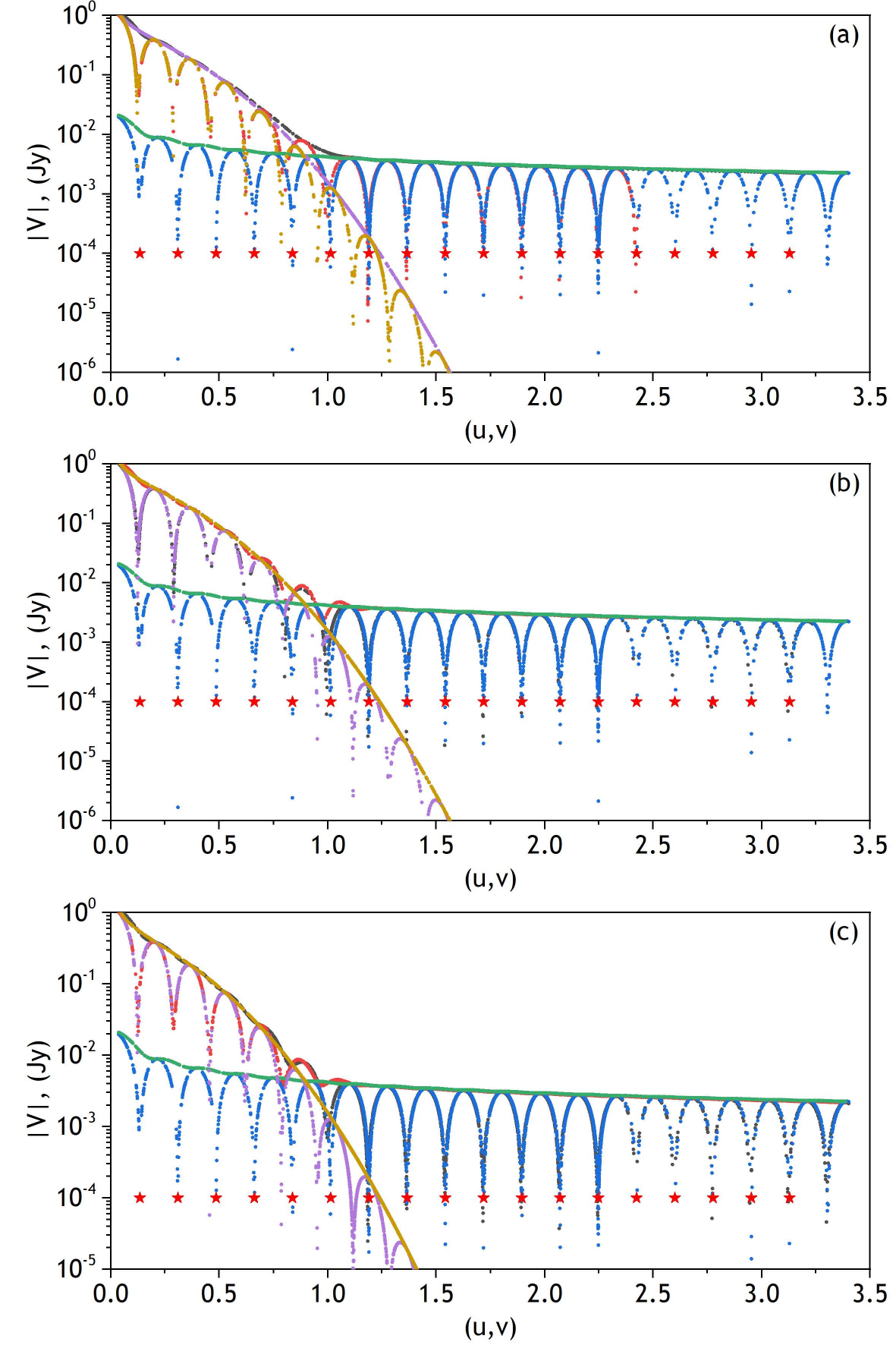}
\caption{Visibility function modulus in two perpendicular directions for the case of one thin ring with radius $r_1=20.0$ and width $w_1=0.1$ $\mu$as and one thick ring with radius $r_2=21.0$ and width $w_2=5.0$ $\mu$as. Numerical observations are shown in red and black. Blue and green are analytical expressions for the thin ring (\ref{Vaxithinring}), purple and yellow are analytical expressions for the approximate thick ring (\ref{Vthick}).}
\label{fig8}
\end{figure}

Let us first consider the simplest case, when the directions to the maximum brightness in both rings coincide, $\phi_1=\phi_2=\pi$. In this case, from the expression (\ref{V1}) it is clear that in the direction $\phi_u=\pi/2$ the imaginary part of the visibility function is zero, i.e. the visibility function must be purely real. Fig.~\ref{fig8} (a) shows the modulus of the visibility function in two perpendicular directions (black and red curves). It is clear from the figure that the visibility function consists of two parts. The flatter visibility function ($u,v>1$ ED) corresponds to the thin ring, and the steeper one ($u,v<1$ ED) to the thick ring. The red asterisks are zeros in the zero-order Bessel function. From the first zero of the Bessel function one can estimate the radius of the thick ring. From the zeros in the region of the thin ring ($u,v>1$ ED) one can estimate the radius of the thin ring. The estimate yields the following values. The first zero is $u_0\approx0.127$ ED, which corresponds to the radius of the thick ring of $r_2\approx21.21$ $\mu$as. For example, the tenth zero is $u_0\approx1.715$ ED, which corresponds to the radius of the thin ring of $r_1\approx20.04$ $\mu$as. It is not possible to determine the width of the thick ring from the envelope zero, since on the scale $u\sim \pi/w$ the thin ring dominates over the thick one in the visibility function. The thickness of the thick ring can be estimated by fitting the envelope model of the thick ring (\ref{Vthick}). Fitting with \texttt{curvefit} yields a value for the thick ring width of $w\approx4.65$ $\mu$as. The blue and green curves show the analytical model of the thin ring (\ref{Vaxithinring}). The purple and yellow curves show the analytical model of the thick ring (\ref{Vthick}).

The case when the directions of maximum brightness in both rings are located in opposite directions ($\phi_1=\pi/2$ and $\phi_2=3\pi/2$) is similar to the previous one. The modulus of the visibility function in two perpendicular directions (black and red curves) is shown in ~Fig.\ref{fig8} (b). The flatter visibility function ($u,v>1$ ED) belongs to the thin ring, and the steeper one ($u,v<1$ ED) to the thick ring. The radius of the rings can be determined by comparing the zeros of the visibility function with the zeros of the Bessel functions. The first zero is $u_0\approx0.126$ ED, which corresponds to the radius of the thick ring equal to $r_2\approx21.37$ $\mu$as. For example, the tenth zero is $u_0\approx1.716$, which corresponds to the radius of the thin ring equal to $r_1\approx20.03$ $\mu$as. The width of the thick ring cannot be determined from the zero envelope, since the thin ring dominates over the thick ring in the visibility function on the scale $u\sim \pi/w$. The thick ring thickness can be estimated by approximating the numerical data with the thick ring model (\ref{Vthick}). Approximation using \texttt{curvefit} yields a value for the thick ring width of $w\approx5.46$ $\mu$as.

It is also necessary to consider the case when the directions of the brightness maxima in the thick and thin rings are located at an angle of 90 degrees to each other $\phi_1=3\pi/4$ and $\phi_2=5\pi/4$. The modulus of the visibility function in two perpendicular directions (black and red curves) is shown in Fig.~\ref{fig8} (c). The flatter visibility function ($u,v>1$ ED) belongs to the thin ring, and the steeper one ($u,v<1$ ED) to the thick ring. Similarly, the radii of the rings are determined by comparing the zeros of the visibility function with the zeros of the Bessel functions. The first zero is $u_0\approx0.128$ ED, which corresponds to the radius of the thick ring equal to $r_2\approx21.04$ $\mu$as. For example, the tenth zero is $u_0\approx1.715$ ED, which corresponds to the radius of the thin ring equal to $r_1\approx20.04$ $\mu$as. It is not possible to determine the width of the thick ring from the envelope zero, since the thin ring dominates over the thick one in the visibility function on the scale $u\sim \pi/w$. The thick ring thickness can be estimated by approximating the numerical data with approximate expressions of the thick ring model (\ref{Vthick}). Approximation using \texttt{curvefit} yields a value for the thick ring width equal to $w\approx5.32$ $\mu$as.

From the above, it can be concluded that the Bessel function zeros can be used to determine the radii of rings quite accurately, whether it is a thin or thick ring. The ring width can be estimated accurately less than $10\%$.

\section{Ring parameter estimates from M87 EHT observations$^{*}$}

Finally, it is worthwhile to consider the application of the ring parameter estimation method using a simplified asymmetric analytical model of a ring with a Gaussian profile to observational data obtained by the Event Horizon Telescope \cite{EHT_2019}. In VLBI observations of M87, the EHT obtained an image containing some ring-like structure with a diameter of 42 $\mu$as and a width of less than 20 $\mu$as (see Table~1 in \cite{EHT_2019}), which they interpreted as the image of a supermassive black hole located in the elliptical galaxy M87.

\begin{figure}
\centering
\includegraphics[scale=0.5]{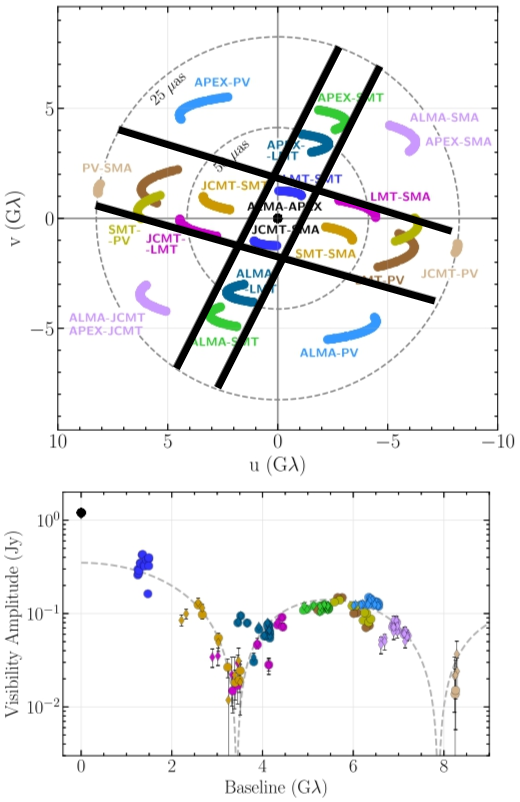}
\caption{Upper panel: $(u,v)$ coverage for M87$^{*}$ EHT observations. Black lines indicate two perpendicular sections that were used to compare the visibility function with theoretical expressions. Lower panel: visibility function measured in M87 $^{*}$ EHT observations. Figure from \cite{EHT_2019}.}
\label{fig3}
\end{figure}

Based on the analysis of the two-dimensional visibility function, we estimate the asymmetry parameter $B$ and the ring diameter (see formula (\ref{axibrightness})), using a simple model of an infinitesimally thin ring. For this purpose, two perpendicular directions of the $(u,v)$ plane occupancy were selected: left-to-right and top-to-bottom (see Fig.~2 of the article \cite{EHT_2019}) presented in the upper panel of Fig.~\ref{fig3}. The selected perpendicular directions are indicated by black solid lines.

\begin{figure}
\centering
\includegraphics[width=0.6\linewidth]{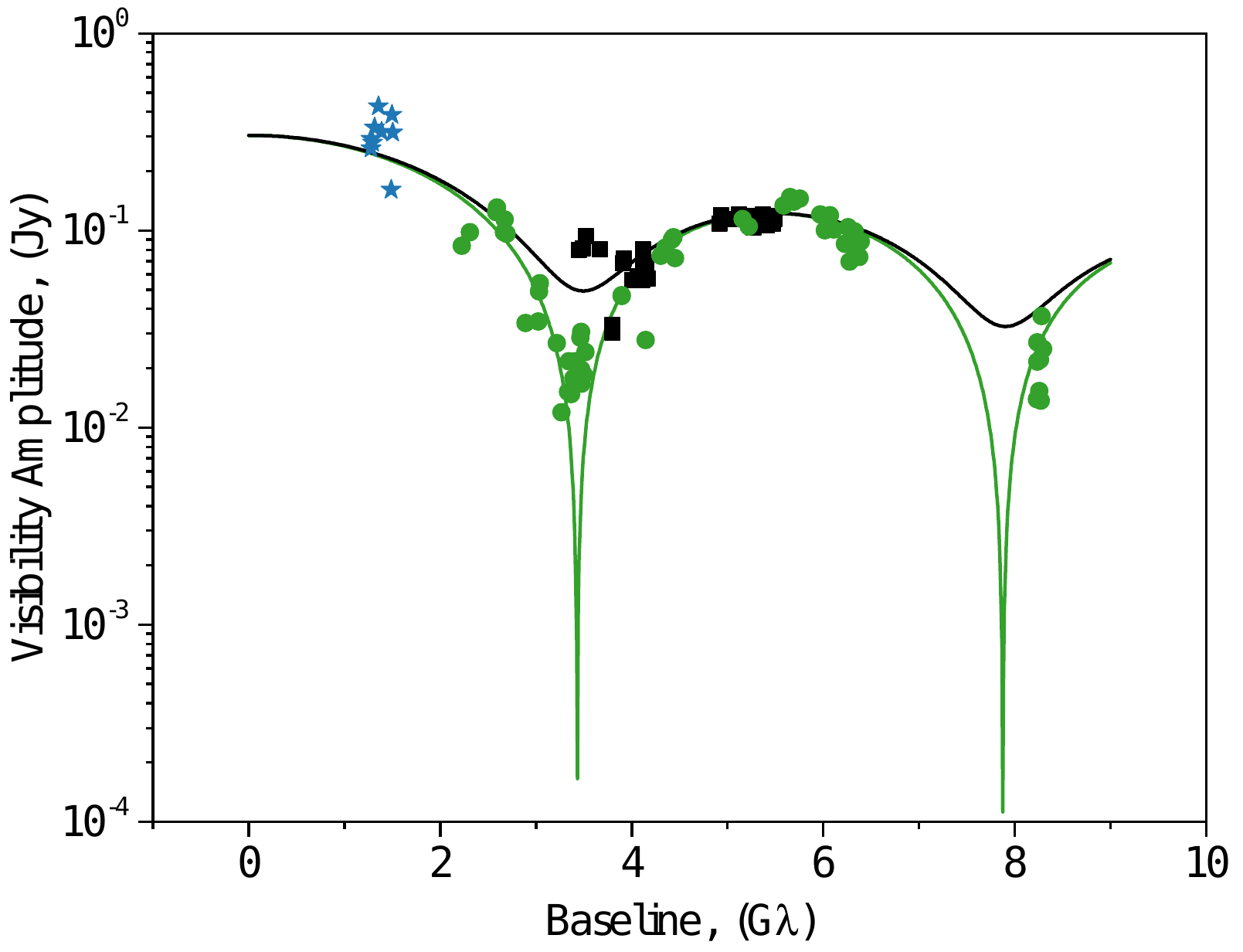}
\caption{Approximation of the visibility function of M87$^{*}$ EHT observations obtained in two perpendicular directions.}
\label{fig4}
\end{figure}

The visibility function for each direction is shown in Fig.~\ref{fig4}. These points were obtained from the visibility function in Fig.~\ref{fig3} (lower panel) using the \texttt{plotdigitizer} program.

The blue dots are common points for the two left-to-right and up-to-down directions and correspond to the blue dots in Fig.~\ref{fig3} (lower panel). The black dots represent the visibility function in the up-down direction, and the green dots represent the visibility function in the left-to-right direction. The green and black curves were obtained by fitting the function (7) in two directions using the \texttt{curvefit} method from the \texttt{scipy} package. The green curve corresponds to the approximation with the angle $\phi_u=\pi/2$, and the black curve, which is the envelope, relates to the approximation in the perpendicular direction with the angle $\phi_u=\pi$. From the first approximation (in the "left-right" direction), the radius of the black hole's shadow was determined to be $r=23.01$ $\mu$as. The second approximation (in the "up-down" direction) allowed us to estimate the asymmetry parameter, which turned out to be $B=0.48$. The ring width could not be determined due to insufficient resolution, since the envelope does not tend to zero. This estimate of the ring radius agrees well with the EHT estimate ($r\approx21$ $\mu$as). The asymmetry parameter was estimated for the first time.

\section{Conclusion}

A simplified asymmetric analytical model of a ring with a Gaussian profile was considered in this paper. For this model, a two-dimensional visibility function was constructed with maximum projections of the interferometer base up to 3.5 Earth diameters. The visibility function was studied in two perpendicular directions. In one of the directions, the visibility function demonstrates rapidly oscillating oscillations, while in the perpendicular direction, a beat is observed. The ring radius and width were estimated from the visibility function zeros in two perpendicular directions. 

By comparing numerical observations of the visibility functions with the position of the Bessel function zeros , it can be concluded that in the case of a thin ring ($w\gg r_0$), even for a sufficiently large zero order (for example, for the tenth zero), its position coincides very well with the visibility function minimum. For a thick ring, the first few zeros (for a width of $w=10$ $\mu$as -- three zeros) coincide quite well with the visibility function minimum. The ring width was estimated from the first zero position in the envelope function. The estimate turned out to be quite rough, but the determination accuracy increases with increasing ring width. And for $w=10$ $\mu$as, the accuracy is less than $5$\%.

In addition, a model consisting of two thin rings was considered, as well as a model of a thin ring embedded in a thick ring. For the first model, the radii of the two thin rings were estimated, and for the second model, the width of the thick ring was also determined. It is shown that analysis of the two-dimensional visibility function in two perpendicular directions allows us to estimate not only the ring radius, but also its width. From a comparison of numerical observations of the visibility function with the position of the zeros of the Bessel function, we can conclude that the radii of the thin and thick rings agree well with the original values. The thick ring width is determined accurately less than $7$\%.

Then, a simplified asymmetric analytical model of the ring with a Gaussian profile was applied to real observations of t M87$^*$, conducted in April 2017 by the Event Horizon Telescope. As a result, the ring radius was estimated to be $r\approx23$ $\mu$as, as well as the asymmetry parameter, $B=0.48$. The ring radius agrees quite well with the EHT data ($r\approx21$ $\mu$as, see Table 1 of \cite{EHT_2019}), and the asymmetry parameter was estimated for the first time.

\section*{Appendix}

In general, the integral for the visibility function (\ref{V1}) with the brightness distribution (\ref{rbrightness}) can be approximated as follows:

\begin{equation}
 V(u,\phi_u)=\pi\int_0^\infty [(2-B)J_0(2\pi ur)-iBJ_1(2\pi ur)\cos(\phi_u-\phi_0)]e^{-(\frac{r-r_0}{\Delta r})^2}rdr
 %=(2-B)A_1-iB\cos(\phi_u-\phi_0)A_2,
 \label{Vapprox}
\end{equation}

To do this, we expand the Bessel functions in a series, assuming that the projection of the base is large enough $2\pi ur\gg1$:

\begin{equation}
    \begin{split}
        &V(u,\phi_u)\approx\frac{2-B}{\sqrt{2u}}\int_0^\infty \sqrt{r}(\sin(2\pi ur)+\cos(2\pi ur))e^{-(\frac{r-r_0}{\Delta r})^2}dr-\\
        &-i\frac{B\cos(\phi_u-\phi_0)}{\sqrt{2u}}\int_0^\infty\sqrt{r}(\sin(2\pi ur)-\cos(2\pi ur))
        e^{-(\frac{r-r_0}{\Delta r})^2}dr.   
    \end{split}
\end{equation}

We will also use the fact that the radius of the ring $r_0$ is much greater than the width of the ring:

\begin{equation}
\sqrt{r}\approx\frac{1}{2}(\sqrt{r_0}+\frac{r}{\sqrt{r_0}}+\ldots)
\end{equation}

As a result,

\begin{equation}
     \begin{split}
         &V(u,\phi_u)\approx\frac{2-B}{\sqrt{8u}}\int_0^\infty (\sqrt{r_0}+\frac{r}{\sqrt{r_0}})(\sin(2\pi ur)+\cos(2\pi ur))e^{-(\frac{r-r_0}{\Delta r})^2}dr-\\
         &-i\frac{B\cos(\phi_u-\phi_0)}{\sqrt{8u}}\int_0^\infty(\sqrt{r_0}+\frac{r}{\sqrt{r_0}})(\sin(2\pi ur)-\cos(2\pi ur))e^{-(\frac{r-r_0}{\Delta r})^2}dr.    
     \end{split}
\end{equation}

The above integrals can be taken explicitly. The solution is written as:

\begin{equation}
 V(u,\phi_u)\approx(2-B)C-iB\cos(\phi_u-\phi_0)D,
 \label{Vthick} 
\end{equation}

where $C=C_1+C_2$, $D=D_1+D_2$ and

\begin{eqnarray}
        &C_1=\sqrt{\frac{\pi r_0}{8u}}\Delta r\frac{1+i}{4}e^{-2\pi ur_0(\frac{\pi u \Delta r^2}{2r_0}+i)}\times\nonumber\\
        &\times\bigg[1+\mathrm{erf}\left(\frac{r_0}{\Delta r}-\pi u\Delta ri\right)-ie^{4i\pi ur_0}          1+\mathrm{erf}\left(\frac{r_0}{\Delta r}+\pi u\Delta ri\right))\bigg]\\
        &C_2=\frac{\Delta r^2}{8\sqrt{8ur_0}}\bigg[\sqrt{\pi}(1-i)(s+2id)e^{-\frac{s^2}{4}-isd} \mathrm{erf}(d-i\frac{s}{2})+\nonumber\\
        &+(1+i)e^{-d^2-ids-\frac{s^2}{4}}\big[2(1-i)e^{ids+\frac{s^2}{4}}+\nonumber\\
        &+\sqrt{\pi}(2d-is)e^{d^2}+\sqrt{\pi}e^{d^2+2ids}(s-2id)(1+\mathrm{erf}(d+i\frac{s}{2}))\big]\bigg]\\
        &D_1=\sqrt{\frac{\pi r_0}{2u}}\Delta r\frac{1+i}{4}e^{-2\pi ur_0(\frac{\pi u \Delta r^2}{2r_0}+i)}\times\nonumber\\
        &\times\bigg[i(1+\mathrm{erf}\left(\frac{r_0}{\Delta r}-\pi u\Delta ri\right))-e^{4i\pi ur_0}(1+\mathrm{erf}\left(\frac{r_0}{\Delta r}+\pi u\Delta ri\right))\bigg]\\
        &D_2=\frac{\Delta r^2}{8\sqrt{8ur_0}}(1+i)\bigg[\sqrt{\pi}(s+2id)e^{-\frac{s^2}{4}-isd} 
        \mathrm{erf}(d-i\frac{s}{2})+e^{-d^2-ids-\frac{s^2}{4}}\big[2(i-1)e^{ids+\frac{s^2}{4}}+\nonumber\\
        &+\sqrt{\pi}(s+2di)e^{d^2}-\sqrt{\pi}e^{d^2+2ids}(2d+is)(1+\mathrm{erf}(d+i\frac{s}{2}))\big]\bigg]
\end{eqnarray}

where $\mathrm{erf}$ is the error function, $s=2\pi u\Delta r$, $d=r_0/\Delta r$.

\bibliography{biblio}% Produces the bibliography via BibTeX.

\end{document}